# Review and Recommendations for using Artificial Intelligence in Intracoronary Optical Coherence Tomography Analysis

Xu Chen[1], Yuan Huang[1,2], Benn Jessney[1], Jason Sangha[1], Sophie Gu[1], Carola-Bibiane Schönlieb[2], Martin Bennett[1,†], and Michael Roberts[1,2,†]

[1]Department of Medicine, University of Cambridge, Cambridge, UK
[2]Department of Applied Mathematics and Theoretical Physics, University of Cambridge, Cambridge, UK
[†]Corresponding authors: mr808@cam.ac.uk, mrb24@cam.ac.uk

## ABSTRACT

Artificial intelligence (AI) methodologies hold great promise for the rapid and accurate diagnosis of coronary artery disease (CAD) from intravascular optical coherent tomography (IVOCT) images. Numerous papers have been published describing AI-based models for different diagnostic tasks, yet it remains unclear which models have potential clinical utility and have been properly validated. This systematic review considered published literature between January 2015 and February 2023 describing AI-based diagnosis of CAD using IVOCT. Our search identified 5,576 studies, with 513 included after initial screening and 35 studies included in the final systematic review after quality screening. Our findings indicate that most of the identified models are not currently suitable for clinical use, primarily due to methodological flaws and underlying biases. To address these issues, we provide recommendations to improve model quality and research practices to enhance the development of clinically useful AI products.

## 1 Introduction

Accurate diagnosis of coronary artery disease (CAD) is essential for clinical assessment and the accurate identification of plaque composition and features could aid in identifying patients at risk of future coronary events. Intravascular Optical Coherence Tomography (IVOCT) offers detailed visualisation of coronary arterial structures,[1,2] with significantly higher resolution than other invasive imaging modalities (such as intravascular ultrasound (IVUS) and angiography)[3] or non-invasive methods (such as magnetic resonance imaging (MRI) and computed tomography (CT)). IVOCT can accurately identify plaque tissue types, (e.g., fibrous tissue, calcium, lipid) and different plaque types (e.g., pathological intimal thickening, fibrocalcific, thin and thick cap fibroatheroma etc.).[4,5] However, OCT datasets are large with structures that are often unclear and variable in appearance, and manual interpretation of hundreds of images per patient is not only time-consuming but also prone to errors[6]. There is a growing demand for automated methods of IVOCT image analysis to address these challenges.

The application of Artificial Intelligence (AI) techniques, particularly Machine Learning (ML), enables rapid image interpretation by automating classification and segmentation processes, thereby potentially accelerating IVOCT analysis **(Figure 1).** However, there is increasing[7–9] concern about the datasets used to train ML models, their validation, reproducibility and generalisability, resulting in a lack of standardisation. We found 172 papers using AI/ML methods have been applied to IVOCT, but there was inconsistent standardisation such as data collection/selection procedures, performance validation and evaluation metrics. While previous reviews offered a broad analysis of AI models for IVOCT analysis,[10,11] we specifically focus on the systematic methodological pitfalls in the current literature. We assess the risk of bias in the literature, incorporating a quality screening stage to ensure that only papers with sufficiently documented methodologies are reviewed. We also provide detailed recommendations across four domains: (1) considerations for collating IVOCT imaging datasets intended for public release; (2) methodological considerations for AI/ML researchers and specific issues regarding validation of results; (3) specific issues regarding the reproducibility of results; and (4) considerations for reviewers conducting peer reviews of manuscripts.

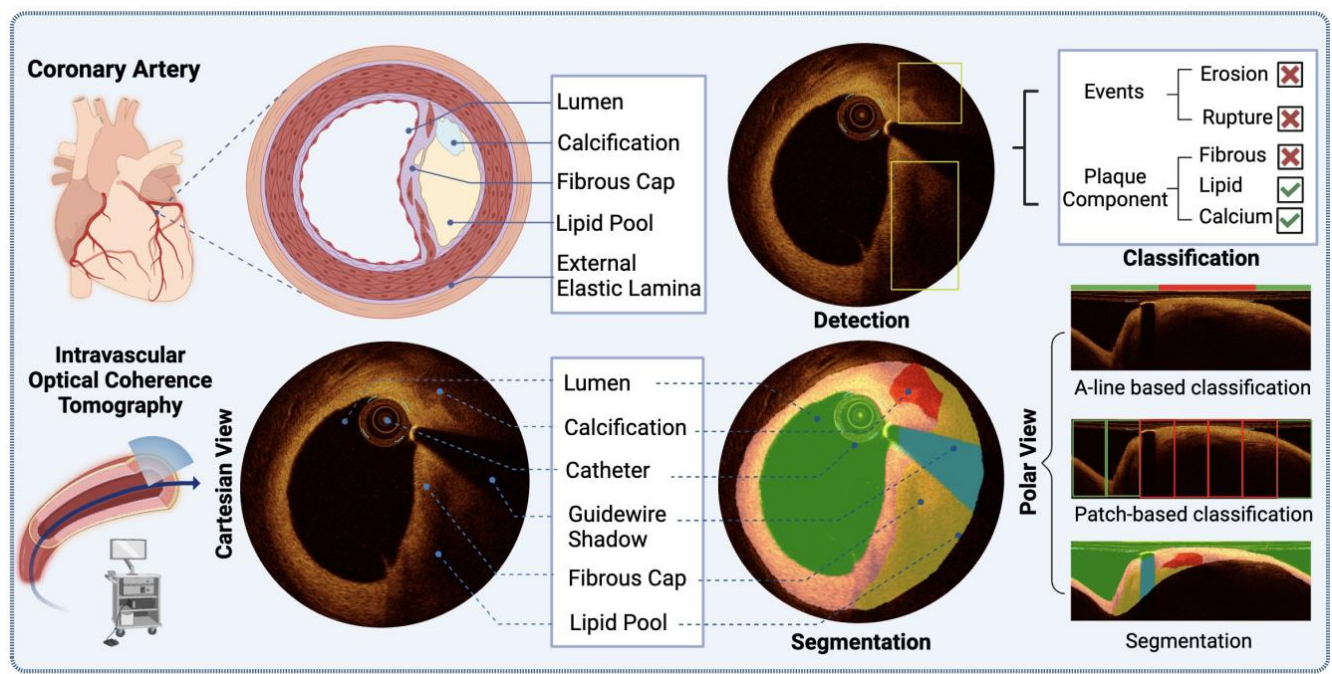

**Figure 1:** An Overview of coronary artery and plaque anatomy, corresponding Cartesian IVOCT image, IVOCT detection and segmentation, and use to predict pathology underlying events, measure plaque components, and classify plaques in the polar view

## 2 Methods

**Review strategy and selection criteria.**

*Initial screening.* To obtain the screening population, we searched SCOPUS for all whose titles or abstracts contained either "OCT" or "optical coherence tomography" along with either "deep learning", "artificial intelligence", "AI", "machine learning", "neural networks", "auto", "ML" or "net"). This search was not case sensitive and returned 5,576 papers whose titles and abstracts were pre-screened to exclude those papers focussed on retinal IVOCT imaging and retain those focussed on coronary arteries (detailed search criteria are detailed in the Supplementary Information).

*Title and abstract screening.* This was performed by three independent reviewers with conflicts resolved via consensus.

*Full-text screening.* Two reviewers performed the full-text screening with conflicts resolved by consensus with a third reviewer.

*Quality screening.* This stage aimed to exclude those deep learning papers which had poor quality documentation of critical details necessary for reproducibility of the method described. We followed the approach in Roberts et al.[9] comparing manuscripts with the Checklist for Artificial Intelligence in Medical Imaging (CLAIM)[12] and excluding those papers which failed any of eight mandatory criteria (detailed in the Supplementary Information). The full CLAIM checklist is reported for each paper which passes the quality screen. Traditional machine learning papers are assessed using the radiomic quality score (RQS)[13] guidelines but no papers were excluded on the basis of their score.

*Assessing the risk of bias in studies.* In order to assess bias in the datasets, predictors, outcomes and model analysis in each paper, we use the PROBAST of Wolff et al.[14] Papers that passed the quality screening stage were split among two reviewers to complete the PROBAST review, with conflicts resolved by a third reviewer.

*Data extraction.* Four reviewers extracted data from the manuscripts, with two reviewers considering each manuscript and resolving conflicts. The full dataset is in the Supplementary Data and forms the basis of this review.

# 3 Results

## 3.1 Study selection

We identified 5,576 papers that satisfied our search criteria and 172 had abstracts or titles relevant to this review, i.e. developing ML methods using IVOCT imaging for diagnostic modelling of CAD. After full-text screening, 71 papers remained and quality screening retained 35/71 papers for consideration in this review **(Table 1)**. Of these, 30/35 were pure deep learning (DL) papers and 3/35 we refer to as traditional ML papers (i.e. non-deep learning ML papers). Two papers developed a hybrid of both approaches [15,16] (**Figure 2**).

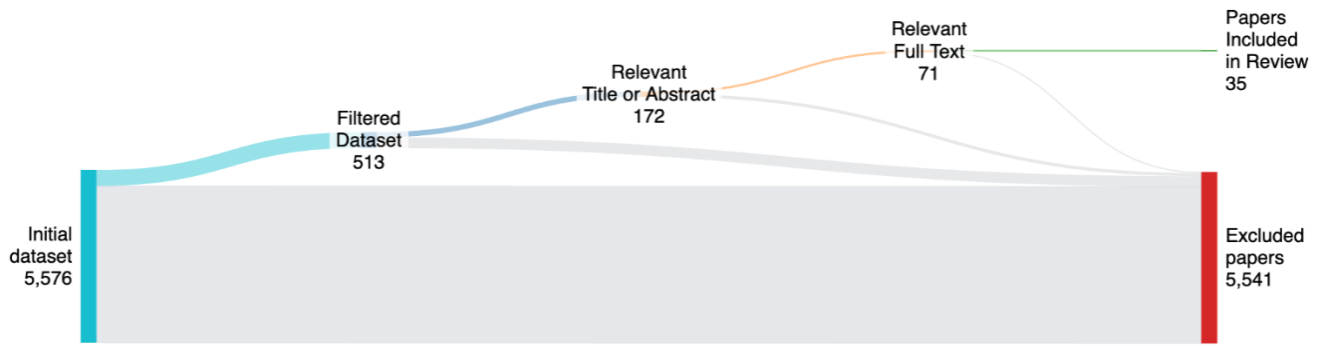

**Figure 2: PRISMA diagram indicating the inclusion/exclusion of papers at each stage of the review.**

*Paper quality screen*

Of the 68 deep learning papers screened using the CLAIM checklist[12], over half (36/68) were excluded for failing mandatory CLAIM checklist criteria (detailed in the Supplementary Information). 19/68 failed just one, 14/68 failed two and 3/68 failed three or more. The two most common reasons for failing the quality screen were insufficient documentation of the data source (28/68) and description of the training approach (13/68). Of 68 papers, 35 passed the initial quality screening, with 30 being DL papers, 3 were traditional ML papers and 2 were a hybrid of these.

In the next stage, DL papers were evaluated using the full CLAIM checklist, while traditional ML papers were assessed with the RQS[13] checklist to establish how adherent the manuscripts were to established guidelines for manuscript completeness.

<u>DL paper quality screen</u>

Only 4/32 papers failed more than 10 items on the 42-point CLAIM checklist and 27/32 failed more than five. The five most common items not satisfied were missing: robustness/sensitivity analysis of models (31/32), details for manual annotation tools (27/32), external model validation (25/32), patient demographics (25/32), and clear details on inclusion/exclusion to obtain patient populations (25/32).

<u>Traditional ML paper *quality screen*</u>

The three papers assessed using RQS received scores of 4,[17] 5,[18] and 8.[19] No papers reported calibration statistics, statistical significance for discrimination statistics, the potential clinical utility, cost-effectiveness nor shared code, data or models. Only one paper performed feature reduction when selecting features[19].

## 3.2 Datasets

**Sources.** To train generalisable and reproducible ML models, it is best practice to train using data from multiple sources, representative of the patient populations under study, and ideally available for other practitioners to access and train/validate against. However, the majority of papers (31/35) used entirely private data sources (one used both a public

**3/13**

and private dataset[20]) and 29/35 used a single data source. Data was primarily from the United States (14/35) and China (13/35) with only 4/35 papers[21–24] using data from multiple countries (**Figure 3**). Only one public dataset was identified

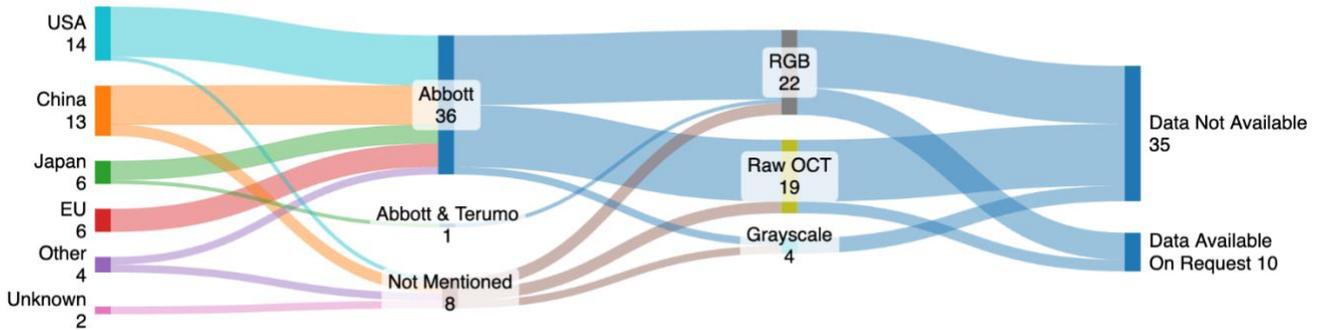

**Figure 3: Overview of data sources and modalities in the literature.**

**Acquisition.** To ensure wider generalisability of ML models, the datasets should be acquired from multiple catheter manufacturers, and the input images processed in the same format. However, only 26/35 papers detailed the catheter used to acquire their IVOCT images, all of which used Abbott manufactured catheters (or prior named corporate entities) for model training. In one paper[24], images acquired using a Terumo catheter are used in the external validation cohort and one paper[20], which uses two datasets, used Abbott catheters for one dataset but did not disclose the manufacturer for the other. In 19/35 papers, the raw catheter output was used to train the models, with 12/35 using derived RGB outputs and 4/35 using grayscale.

**Ground-truth annotation.** Annotation of ground truth in IVOCT images, often by trained experts using specific software, is critical for model training and subsequent validation. However, 7/35 papers[17,19,23,25–28] did not detail who performed manual annotation of ground-truth data, while 8/35 reported one annotator was involved,[15,18,21,29–33] 16/35 reported two and 4/35 reported using more than three annotators.[24,34–36] Only 10/35 papers provided details for the specific annotation tools used, namely Abbott's offline review software[21], MATLAB's Video Labelling Tool, [17] Amira,[15,37] ImageJ[33], OCTOPUS[29,38,39] and LabelMe.[32,40]

**Cohort selection.** AI is being proposed for identification of native atherosclerosis, but also for stent optimisation and complications of PCI, and ideally any model should be generalisable across different types of CAD and vessels. However, only 15/35 papers reported their study inclusion criteria and only 12/35 papers specified exclusion criteria. One paper disclosed exclusion criteria for one of their datasets but not the other.[20]

**Sample sizes.** Coronary IVOCT pullbacks may be obtained from the same or different arteries from the same patient and on multiple occasions. However, datasets for AI models should comprise large numbers of independent pullbacks, and generalisability/reproducibility claims should not be made based on performance on highly selected IVOCT frames, for example those with 'classical' architecture, known measurements, and an absence of imaging artefacts, which may not be reflective of real-world clinical practice. Manuscripts which split at the patient, pullback, lesion and frame level considered a median of 64, 391, 52 and 55 patients respectively although the number of patients is not mentioned in 5/13 models which split at lesion-level and 5/7 manuscripts which split at frame-level. Only 5/35[20,21,24,34,41] papers had a dataset consisting of more than 100 patients. Surprisingly, 29/35 papers used only selected frames from pullbacks in their model development, rather than the whole pullback, with no documentation for how these subsets of frames were selected and real-world applicability is consequently limited.

### 3.3 Model Development

**Outcomes of interest.** The power of AI to identify and measure tissue types on IVOCT, and thus classify plaques, depends upon accurate localisation of tissues with a range of different appearances, using dense segmentations, bounding boxes or A-line- or patch-based classification (**Figure 1**). 17/35 papers described segmentation models for calcium (12/17), lipid (5/17), or fibrous (4/17) plaques, and 10/35 papers described methods which identified a bounding box around tissues including calcified (1/10) or mixed plaques (7/10) and thin fibrous caps (2/10). Classification models were developed in 8/35 papers, focussed on classifying entire frames as showing lipid and calcium plaques (6/8), rupture[22] (1/8) and thin cap fibroatheromas (1/8)[20]. Three of these papers describe classification models operating on the A-line raw IVOCT data and aimed to distinguish lipid vs calcium.[19,23,42] (**Figure 4**).

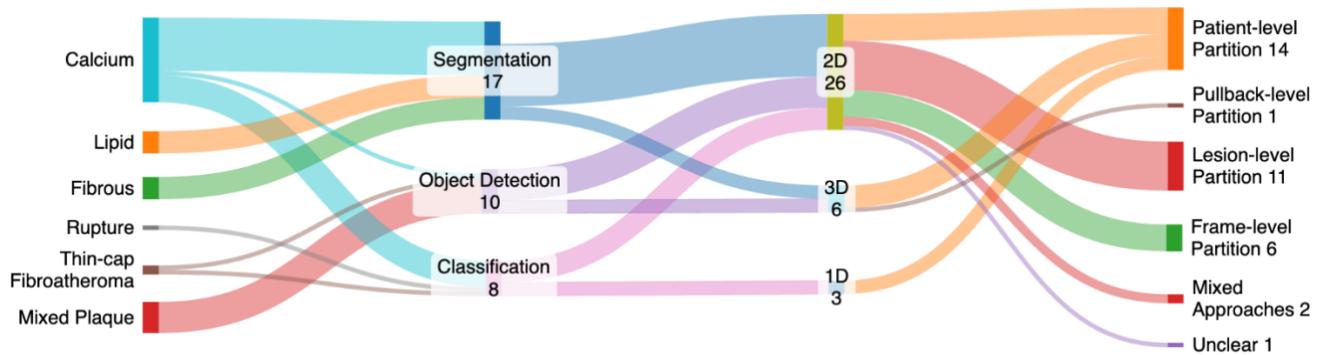

**Figure 4: Overview of model development strategies in the literature.**

**Dataset partitioning.** Only 14/35 papers partitioned their datasets at the patient-level for their model development and evaluation, whereas 1/35 partitioned[24] at pullback-level, 11/35 at the lesion-level and 6/35 at the frame-level. One paper[43] trained both a classifier and segmentation models, using a patient-level split and lesion-level split respectively. One paper[20] trained a model independently for two datasets, with a lesion-level split for one and an unclear approach for the other. Finally, one paper was unclear in how the data had been split [44]. Failing to split data at the patient-level risks data from the same patient, pullback or lesions being used for both training and model evaluation and the likelihood of optimistic performance reporting. Most papers (19/35) used cross-validation to develop their models, with 8/35 using a fixed internal validation set and 6/35 relying only on internal holdout data. One paper[32] was unclear in how the model was evaluated and another[20] developed two models, one using cross-validation and another using a fixed internal validation cohort.

**Data pre-processing.** Despite raw IVOCT imaging data being acquired in polar coordinates, most papers (18/35) performed their analysis on Cartesian transformed images, which are more familiar to clinicians and one used both polar and Cartesian images[20] **(Figure 1)**. Real-world IVOCT pullbacks contain multiple artefacts[45], a significant proportion of which affect interpretation of the underlying vessel wall. However, only 12/35 papers applied denoising techniques to suppress these artefacts (e.g., using Gaussian and Gabor filters). Segmentation quality and measurements can be greatly affected by image resolution and size, both of which are commonly standardised during pre-processing. The resampled image size is mentioned in only 21/35 papers with only two papers considering an image resolution above 512 x 512 pixels [23,24].

**Model inputs.** IVOCT images are acquired using a spinning catheter sampling radially as it is pulled along the artery, these radial samples are stacked to form the 2D frames of the final image. The 2D frames themselves are stacked to form a 3D volume. A majority of papers (26/35) developed their models using 2D frames as inputs with 6/35 considering the pullback as 3D volumetric data[15,21,24,30,37,46] and 3/35[19,23,42] focussing on the 1D A-line data. None of the papers consider the spiral nature of IVOCT data acquisition in their modelling approach.

**Model architecture.** There is a rapidly increasing selection of available AI applications and model architectures reflected in the published literature**.** Of the 17 papers describing segmentation models, 14/17 built upon existing DL architectures including: U-Net-like (5/14),[16,24,37,40,47] SegNet (4/14),[32,33,48,49] Deeplabv3+ (4/14),[29,38,39,48] and Transformer-based models (1/14),[25] while the other DL-based paper trained a custom architectures. The 2/23 papers which focussed on traditional ML algorithms evaluated a random forest[17], XGBoost[17] and support vector machines[18] for pixel-level classification. For the 10/35 papers describing object detection applications, all used DL with 2/10 papers using a Faster R-CNN[26,31] and ResNet,[28,30] while the rest employed a highly diverse range of architectures including a vision transformer,[21] DenseNet,[41] Inception-V3,[34]

EfficientNet,[34] Mask RCNN,[15] VGG[36] and autoencoders[27]. For the eight classification models, 3/8 papers described DL models established architectures,[22,23,35] 4/8 used custom architectures[20,42,44,50] and one[19] used a support vector machine and random forest for A-line classification.

**Metrics.** In addition to different architectures, IVOCT segmentation performance was described using several different metrics. The Dice score was most frequently used (11/17) to assess IVOCT segmentation model performance, while pixel-based accuracy/sensitivity, Intersection Over Union or F1-score were used in other papers. Classification model performance was assessed using a diverse range of metrics, namely recall/sensitivity, precision (PPV), specificity, accuracy, F1 score, NPV, the area under the receiver operating characteristic curve and Bland-Altmann agreement. Object detection models used many of the classification metrics and the overlap rate, mean average precision and Brier score.

### 3.4 Risk of bias assessment

AI models can be very prone to bias, most commonly arising from the participants, predictors, outcome and analysis (see Supplementary Data), The risk of bias (ROB) and concerns of applicability were assessed for all papers using the prediction model risk of bias assessment tool (PROBAST) guidance. We found a high ROB in at least one of the four domains in 20/35 papers (see Supplementary Data).

**Participants.** The ROB was rated as low for the participants domain in only 5/35 papers,[21,22,24,34,41] with a high ROB found in 17/35 papers (see Table 1). This was primarily due to the selection of samples at region-, segment- or frame-level, with a risk of validating the model using patients that the model was trained upon. Risks of bias were also increased due to: selecting a subset of pullback frames without clear inclusion/exclusion criteria; using only well-defined samples and dropping those with equivocal findings; only sampling from selected plaque pathologies; using a data subset where the demographics varied significantly from the full cohort; and only using pullbacks from patient groups where particular plaque compositions were significantly enriched. The ROB was rated as unclear for the Participants domain in 13/35 papers in which insufficient demographics and recruitment information were given, and was found for both private and public datasets.

**Predictors.** Almost all papers (32/35) developed DL models in which the predictors are abstract and unknown imaging features. Therefore, the ROB was rated as unclear for these papers. The remaining three papers[17–19] were found to have a low ROB, relying on pre-defined hand-engineered features derived from the images.

**Outcome.** All papers included in this review described models for classifying plaque types or localising their components with these outcomes defined using consensus recommendations.[6] Therefore, all papers were rated low ROB in this domain.

**Analysis.** Most papers (23/35) were found to have a low ROB for their analysis (see Table 1). High ROB was found in 9/35 papers due to small sample sizes[17,23,27,32,33,39,40,47] or inappropriate performance evaluation.[31] Two papers had an unclear ROB[30,49] as they did not report the proportion of positive samples in their dataset.

### 3.5 Code and model availability
AI model performance can vary greatly between centres and datasets, and external reproduction of results is required to be confident of their generalisability. However, no papers provided detailed instructions or open-source code to allow for the external reproduction of results. Most papers also did not share any data (28/35), with the remainder stating that data was available on reasonable request. Many papers did not mention the software in which their models were implemented (13/35) and of those which did, most used MATLAB (12/35) or Python (10/35). One paper used a commercial tool OctPlus[24].

## 4 Discussion
IVOCT is the highest-resolution widely available modality for the imaging of the coronary arteries and the only one able to identify and measure high-risk thin fibrous caps. In 2024, European guidelines were updated to strongly recommend the use of IVOCT to guide stenting of complex lesions. However, accurate interpretation of IVOCT imaging requires significant training and is a barrier to the scaling of IVOCT adoption. AI models hold significant promise for far faster and more scalable analysis compared with a human reader. Therefore, with increasing availability of IVOCT data and the ability to analyse them with deep learning tools and hardware, it is only natural that we see increasing appetite from researchers to develop AI models for interpreting IVOCT images. However, we found that although published studies show considerable promise and potential in this field, many are burdened with methodological and reporting deficiencies, with most of the reviewed literature not ready for clinical application. We have identified issues around dataset documentation, methodologies, reproducibility and biases in study design, which we now summarise and suggest recommendations to improve the evidence base to allow for wider adoption of AI tools for automated IVOCT interpretation.

In general, there is a strong preference in the literature towards training of deep learning models rather than more traditional radiomics approaches, with only three focussing on the latter (only one since 2020). Additionally, very few papers developed models to give a frame-level classification for disease status, with all others focussed on localising the disease through segmentation and object detection within the image. All the papers which disclosed their catheter manufacturer reported using images acquired by Abbott catheters for model training, likely due to data availability with Abbott distributing the majority of IVOCT catheters globally. Consequently, the majority of models are likely applicable only to Abbott acquired imaging.

**Image acquisition and datasets.**

*Spiral acquisitions.* IVOCT images are acquired by a spinning catheter pulled down an artery, with image data acquired as 1D profiles (A-lines) in a spiral. These 1D grayscale acquisitions are often stacked into (artificially RGB coloured) 2D frames, which themselves are then stacked into 3D images. No papers in this review considered the spiral nature of the acquisition in their modelling, rather focussing the IVOCT images as 2D/3D acquisitions. It is likely that incorporating the acquisition technique into the methodology will improve model quality by reducing known artefacts (such as seam artefacts) that occur due to stacking into 2D frames.

*Dataset sizes.* We found that many papers used relatively small IVOCT datasets for model development with a median of 55 patients and only six studies using datasets with more than 100 patients. Developing models using small-scale datasets risks introducing potential biases into the model, limiting its generalisability, and findings should be approached cautiously. For deep learning models, it is common to require many thousands of training samples due to their over parametrisation, which is more achievable at the 2D frame level but often hard to achieve (and of unclear necessity) at the 3D pullback level.

*Dataset diversity.* Given the real-world diversity of IVOCT imaging data collected from patients of different backgrounds, it is unfortunate that only six studies utilised data from multiple countries with others training primarily with data from the United States and China. This bias in geographic scope will likely limit model applicability, and it is of primary importance to ensure a diversity of patients and of disease profiles. to avoid bias in the model and enhance its applicability.

*Ground truth.* Clinicians are trained to interpret the RGB cartesian IVOCT images, derived from the raw grayscale polar IVOCT data. Surprisingly, however, in around half of papers it is the raw image data that is used to train models and it is generally unclear how the ground truth segmentation labels were generated when these data are never observed in clinical practice.

*Inclusion/exclusion criteria.* Reporting inclusion and exclusion criteria is crucial for understanding a model's training population and likely limitations of applicability. Their absence was particularly striking in our review, with only fifteen papers providing detailed inclusion criteria and exclusion criteria reported in only twelve papers. The inclusion and exclusion criteria varied widely between studies (depending on area of focus), highlighting the heterogeneity in patient populations used for training and underscores the need for transparent reporting to allow for fair study comparisons and improve reproducibility.

**Methodologies for model building**

*Dataset partitioning.* Most papers did not partition their datasets at the patient level and used only an undisclosed subset of frames from the pullback for training and evaluation. This leads to a high-risk of data leakage in the literature whereby frames from the same patient are being allocated to train/validation/holdout cohorts, and similar plaque features from adjacent frames may inadvertently influence model performance and lead to optimistic performance.

*Validation issues.* For model development, twice the number of papers employed cross-validation against those using a single fixed validation cohort to evaluate their model performance. Most models developed using a single private dataset and external validation is absent in most papers, both of which raise important concerns around the generalisability of the models to new populations. Furthermore, with the widespread lack of disclosure of inclusion/exclusion criteria and demographic information on the training population, our review of bias assessment gives a high concern in the ability of the models to generalise to new populations for a majority of papers.

**Reproducibility of the existing literature**

*Manuscript documentation.* In general, the literature is poorly documented for reproducibility, as shown by the quality screening, which removed over half of all papers considered. This was primarily due to insufficient documentation of data sources and model training descriptions. However, as discussed previously, the provenance of data, the inclusion/exclusion criteria and the demographics of the training cohort are critical for accurately understanding the cohort that the results may reproduce. In addition, for those models developed and assessed against selected frames from pullbacks, rather than full images themselves, it is hard to reproduce this selection without a sufficient description of the process. Similarly, without sufficient descriptions of the model training procedure, it is impossible to reproduce the results of any experiments.

*Open science.* The only public dataset that is considered in the literature, CCCV-IVOCT from 2017, appears no longer to be available or accessible. This is very unfortunate as it prevents the transparent benchmarking of model performance and is in contrast with many application areas of AI. For example, there are several large (curated) datasets publicly available for AI model development using Chest X-Rays[51,52], and the lack of this resource may be due to IVOCT being perceived as a niche modality. Similarly, for AI methods in other clinical domains it is common for academic papers to openly share code and trained models to allow for an easier assessment of both whether the proposed model outperforms prior models and how well models generalise to new data. Without such convenience, each author must develop and validate their tools in isolation without performance benchmarks to compare against. However, we find that no papers share their code nor were trained models shared.

**Recommendations**

| |
|---|
| *Validation strategies.* Before developing models, developers must have a clear validation strategy with external datasets held out during model development for assessing the generalisability of the models. |
| *More open science*. The IVOCT research community should aim to publicly release (and permanently archive) more IVOCT datasets from different demographics, to allow for better (transparent) benchmarking of model performances. Progress by the research community would be accelerated by a more transparent and open sharing of code and models to allow for easier assessment of how well models generalise to new data. Platforms such as Hugging Face and Code Ocean allow for rapid deployment of models reproducibly to a wide audience of model developers. |
| *Encourage full pullbacks use.* Models should be trained and evaluated against full IVOCT pullbacks. If only selected frames are used in model training, the applicability will likely be limited in the real world and model performance will likely be optimistic. If full pullbacks are not used, it should be clearly stated as a limitation of the study. |
| *Dataset partitioning.* Dataset sizes should be primarily reported at the patient level and partitioning into training, internal validation and holdout cohorts should be performed at that level. |
| *Dataset sourcing.* Standardised reporting guidelines should be adopted to ensure transparency regarding dataset origins and patient demographics. Authors should adhere to established guidelines for reporting inclusion and exclusion criteria to ensure transparency and consistency in reporting. Model developers should avoid building models using collections of |

> 2D images which are of unknown provenance, as the inherent structural relationships between images from the same pullback give rise to risks of reporting overly optimistic performance.

> *Ground truth.* Manuscripts should clearly disclose how the images were assigned their ground truth labels. For segmentation models this should include information on whether the false-colour RGB images were those labelled and the software used. The known performance differences between experienced and junior IVOCT readers[53] means that it is important to disclose the annotator's experience level. For classification models, if disease labels (e.g. PIT, AIT and TCFA) are assigned based on the IVOCT images themselves this risks incorporation bias (where labels are not independent of the predictors), therefore authors should also compare model performance to labels assigned from other sources e.g. histopathology slices.

> *Improving documentation.* During manuscript drafting, we recommend that authors assess their papers against established standards such as CLAIM[12], RQS[13], PROBAST[14], REFORMS (Reporting Standards for Machine Learning Based Science)[54], TRIPOD (Transparent Reporting of a Multivariable Prediction Model for Individual Prognosis or Diagnosis)[55] and QUADAS (Quality Assessment of Diagnostic Accuracy Studies)[56]. For manuscript reviewers and journal editors, we recommend using these checklists to identify weaknesses in methodology reporting when giving feedback on manuscripts.

## 5 Conclusions

In this systematic review, we have examined the published literature on AI/ML methodologies applied to the diagnosis of CAD using IVOCT, focusing on the quality, reproducibility and potential clinical utility of these methodologies. In their current reported forms, few of the AI-based models reviewed are immediately suitable candidates for clinical translation for the diagnosis of CAD. To enhance the likelihood of these models being integrated into future clinical trials, we recommend (1) using datasets with precise and transparent descriptions of data collection, pre-processing, and any transformations applied; (2) providing thoroughly documented manuscripts with detailed methodologies to ensure that studies can be accurately replicated; (3) Conducting comprehensive external validation with independent datasets to ensure the model's performance is reliable and generalizable across different populations and clinical settings.

## Sources of Funding

This work was supported by British Heart Foundation Grants FS/19/66/34658, RG71070, RG84554, BHF Cambridge Centre for Research Excellence, EPSRC Cambridge Maths in Healthcare Centre Nr. EP/N014588/1, Cambridge NIHR Biomedical Research Centre and National Institutes of Health R01 HL150608.

## Disclosures

MB and MR are founders of Octiocor Ltd., a company developing software for analysing IVOCT imaging.

**Table 1.** Key data extracted for each paper (abbreviations: Obj, Object; AUS, Australia; BE, Belgium; BR, Brazil; CN, China; ES, Spain; GER, Germany; GR, Greece; IL, Israel; IT, Italy; JP, Japan; KR, South Korea; MY, Malaysia; USA, United States of America).

| Authors | Model Type | Data Source | Data Format | Split Level | Patients | PROBAST Participants | PROBAST Analysis |
|---|---|---|---|---|---|---|---|
| Araki et al.[21] | Obj. Detection | JP, USA, CN, BE | RGB | Patient | 581 | Unclear | High |
| Avital et al.[47] | Segmentation | IL | RGB | Frame | Unclear | Low | Low |
| Cheimariotis et al.[23] | Classification | GR, JP | RGB | Patient | 33 | High | Low |
| Chu et al.[24] | Segmentation | AUS, USA, JP, ES, CN | RGB | Pullback | 391 | Low | Low |
| Gharaibeh et al.[49] | Segmentation | USA | Raw | Lesion | Unclear | Unclear | High |
| He et al.[30] | Obj. Detection | CN | Raw | Patient | 18 | Low | Low |
| Holmberg et al.[40] | Segmentation | GER | RGB | Patient | 51 | High | Low |
| Kolluru et al.[16] | Classification | USA | Raw | Lesion | Unclear | Unclear | Low |
| Kolluru et al.[42] | Segmentation | Unclear | Raw | Patient | 48 | High | Low |
| Lee et al.[29] | Classification | USA | Raw | Lesion | 79 | Low | Low |
| Lee et al.[50] | Classification | USA | Raw | Lesion | 49 | Low | Low |
| Lee et al.[43] | Segmentation | IT | Raw | Hybrid | 68 | High | Low |
| Lee et al.[44] | Segmentation | USA | Raw | Unclear | Unclear | High | Low |
| Lee et al.[48] | Segmentation | USA | Raw | Lesion | 55 | High | High |
| Lee et al.[38] | Segmentation | USA | Raw | Lesion | 41 | High | Low |
| Lee et al.[39] | Segmentation | USA | Raw | Frame | 41 | High | High |
| Li et al.[37] | Segmentation | CN | Raw | Patient | 45 | Unclear | Low |
| Liu et al.[25] | Classification | CN | Grayscale | Patient | 60 | Unclear | Low |
| Liu et al.[36] | Obj. Detection | CN | Raw | Frame | Unclear | High | Low |
| Liu et al.[20] | Segmentation | Unclear | RGB | Hybrid | 540 | High | Low |
| Min et al.[41] | Obj. Detection | KR | Raw | Patient | 667 | Unclear | High |
| Niioka et al.[34] | Obj. Detection | JP | Raw | Patient | 1791 | Unclear | Low |
| Oliveira et al.[33] | Segmentation | BR | Grayscale | Patient | 51 | Unclear | Low |
| Park et al.[22] | Classification | JP, USA, CN, BE | RGB | Patient | 581 | Unclear | Low |
| Prabhu et al.[19] | Classification | USA | Raw | Patient | 49 | High | High |
| Rajkumar et al.[17] | Segmentation | MY | RGB | Frame | Unclear | High | High |
| Ren et al.[32] | Segmentation | JP | RGB | Lesion | 3 | High | High |
| Roy et al.[27] | Obj. Detection | USA | Raw | Lesion | Unclear | Unclear | Low |
| Shalev et al.[18] | Segmentation | USA | Raw | Lesion | Unclear | Unclear | Low |
| Shi et al.[28] | Obj. Detection | CN | Grayscale | Lesion | Unclear | High | Unclear |
| Sun et al.[15] | Obj. Detection | CN | Grayscale | Patient | 83 | Unclear | Low |
| Sun et al.[26] | Obj. Detection | CN | Raw | Frame | Unclear | High | Unclear |
| Sun et al.[31] | Obj. Detection | CN | RGB | Frame | Unclear | Unclear | Low |
| Wu et al.[46] | Segmentation | CN | RGB | Patient | 70 | High | Low |
| Yin et al.[35] | Classification | CN | RGB | Lesion | 31 | High | High |